\documentclass[conference]{IEEEtran}
\IEEEoverridecommandlockouts
\usepackage{cite}
\usepackage{amsmath,amssymb,amsfonts}
\usepackage{algorithmic}
\usepackage{graphicx}
\usepackage{textcomp}
\usepackage{pgfplots}
\pgfplotsset{compat=1.18} 
\usepackage{colortbl}
\usepackage{url}
\usepackage{siunitx}
\usepackage{tabularx}
\usepackage{booktabs}
\usepackage[T1]{fontenc} 
\usepackage[normalem]{ulem}
\def\BibTeX{{\rm B\kern-.05em{\sc i\kern-.025em b}\kern-.08em
    T\kern-.1667em\lower.7ex\hbox{E}\kern-.125emX}}

\usepackage{listings}

\colorlet{punct}{red!60!black}
\definecolor{background}{HTML}{EEEEEE}
\definecolor{delim}{RGB}{20,105,176}
\colorlet{numb}{magenta!60!black}

\lstdefinelanguage{json}{
    basicstyle=\normalfont\ttfamily,
    numberstyle=\scriptsize,
    stepnumber=1,
    numbersep=8pt,
    showstringspaces=false,
    breaklines=true,
    frame=lines,
    backgroundcolor=\color{background},
    literate=
     *{0}{{{\color{numb}0}}}{1}
      {1}{{{\color{numb}1}}}{1}
      {2}{{{\color{numb}2}}}{1}
      {3}{{{\color{numb}3}}}{1}
      {4}{{{\color{numb}4}}}{1}
      {5}{{{\color{numb}5}}}{1}
      {6}{{{\color{numb}6}}}{1}
      {7}{{{\color{numb}7}}}{1}
      {8}{{{\color{numb}8}}}{1}
      {9}{{{\color{numb}9}}}{1}
      {:}{{{\color{punct}{:}}}}{1}
      {,}{{{\color{punct}{,}}}}{1}
      {\{}{{{\color{delim}{\{}}}}{1}
      {\}}{{{\color{delim}{\}}}}}{1}
      {[}{{{\color{delim}{[}}}}{1}
      {]}{{{\color{delim}{]}}}}{1},
}

\definecolor{success}{HTML}{98B4D4}
\definecolor{fail}{HTML}{E15D44}

\begin{document}

\title{Intent-Based Network for RAN Management with Large Language Models\\
}

\author{\IEEEauthorblockN{
        Fransiscus Asisi Bimo\IEEEauthorrefmark{1},
        Maria Amparo Canaveras Galdon\IEEEauthorrefmark{2,3},
        Chun-Kai Lai\IEEEauthorrefmark{1},\\
        Ray-Guang Cheng\IEEEauthorrefmark{1},
        and
        Edwin K. P. Chong\IEEEauthorrefmark{3}
       \\
  }
\IEEEauthorblockA{\IEEEauthorrefmark{1}
National Taiwan University of Science and Technology, Taiwan
  }
\IEEEauthorblockA{\IEEEauthorrefmark{2}
NVIDIA, USA
  }
\IEEEauthorblockA{\IEEEauthorrefmark{3}
Colorado State University, USA  \\
  }
  Email: crg@mail.ntust.edu.tw
}

\maketitle

\begin{abstract}
Advanced intelligent automation becomes an important feature to deal with the increased complexity in managing wireless networks. This paper proposes a novel intent-based automation for Radio Access Network (RAN) management, leveraging Large Language Models (LLMs). The proposed method enhances intent translation by autonomously interpreting high-level objectives, reasoning over complex network states, and generating precise RAN configurations through an LLM-integrated agentic architecture. We propose a structured prompt engineering technique and demonstrate that the network can automatically improve its energy efficiency by dynamically optimizing critical RAN parameters through a closed-loop mechanism. It showcases the potential for robust resource management in RAN by adapting strategies based on real-time feedback via LLM-orchestrated agentic systems.

\end{abstract}


\section{Introduction}

The rapid increase in diverse service types and dynamic resource demands in RAN introduces unprecedented operational complexity. Even though Open RAN enables the integration of components from multiple vendors, it can also increase configuration complexity and the risks of misconfiguration. This configuration heavily relies on low-level, manual procedures, which makes agility and scalability difficult to achieve \cite{Azariah_2024}.

Intent-Based Network (IBN) has emerged to manage networks by simplifying detailed technical configurations with minimal external intervention \cite{9925251}. An intent serves as a high-level, human-expressible declaration of network goals, rather than specified methods for achievement. It acts as a structured set of expectations, encompassing requirements, goals, conditions, and constraints \cite{intent3gpp}. An IBN then implements and manages these intents \cite{rfc9315}, fundamentally abstracting network complexity and allowing users and customers to request services without requiring detailed knowledge of their underlying provision. It is crucial to note that "intent" is a fundamental term with various interpretations across key Internet and Telco bodies. The Internet Research Task Force (IRTF) in RFC 9315 \cite{rfc9315} defines intent as a high-level, declarative specification of network goals and desired outcomes, emphasizing it should be network-wide, outcome-driven, and vendor-agnostic. Similarly, 3GPP, in its standardization for 5G networks, defines intent as a set of expectations including requirements, goals, conditions, and constraints, focusing on "what" needs to be achieved \cite{intent3gpp}. They further categorize intents based on roles such as Intent-CSC, Intent-CSP, and Intent-NOP. TM Forum's IG1230 document \cite{tmforum} concurs, defining Intent as "the formal specification of the expectations, including requirements, goals, and constraints, given to a technical system." This diversity underscores the importance of clear intent translation.

Although IBN offers significant potential, a major challenge lies in accurately and dynamically translating high-level user intents into precise RAN instructions. For IBN to work well and for networks to become fully autonomous, how accurately and reliably intent is translated is crucial. Understanding intents, which use natural language, requires intelligent translation. Incorrect translations can result in misconfigurations and severe network failures \cite{ntia_report_2023}.

The advent of LLMs presents a powerful opportunity to overcome this critical translation gap. Some popular LLMs demonstrate excellent performance in processing text input, excelling in their ability to understand and generate natural language text \cite{llama, meta}. This makes them ideal as intelligent intermediaries capable of interpreting formalized intents and generating highly granular RAN configurations. Their advanced semantic understanding allows them to capture the subtle nuances and context inherent in complex service requirements, a significant leap beyond traditional rule-based or statistical methods.

The close loop mechanism of an IBN system, as defined in \cite{9925251}, comprises five components: Intent Profiling, Intent Translation, Intent Resolution, Intent Activation, and Intent Assurance. The interaction of these five components enables the IBN system to provide autonomous closed-loop operations. Intent Translation becomes the most critical and complex component. The translation process requires not only converting syntax but also understanding the meaning to accurately capture a user's intent. Misinterpretations at this stage can lead to inefficient resource utilization, poor Quality of Service (QoS), or even network instability. The effectiveness of an intent-based RAN system is therefore fundamentally constrained by the accuracy and adaptability of its intent translation capability.

Existing studies on closed-loop network management, such as \cite{work1}, translate intent using catalog databases. These approaches lack the advanced semantic understanding offered by LLMs. Separately, work in \cite{work2} uses Generative AI for intent management. However, it misses a crucial closed-loop mechanism, preventing continuous self-adaptation. The survey by \cite{10207407} highlights key challenges in IBN, specifically concerning its goal to enhance automation and simplify network configurations. A major hurdle is the absence of a formal definition for "intent," primarily because various stakeholders hold differing views on its significance. Crucially, this underscores the importance of intent translation. It is the essential step for an IBN system to accurately interpret users' high-level goals and requirements into actionable network policies.

This paper introduces an Intent-Based Wireless Network for RAN configuration that integrates LLMs as a core component to enhance intent translation. We elaborate on a structured prompt technique that leverages the cutting-edge semantic capabilities of LLMs. We demonstrate the system’s capability to dynamically optimize RAN parameters in an energy efficiency use case through a closed-loop mechanism.
The manuscript is organized as follows. Section \ref{Proposed Architecture} discusses the proposed architecture. Section \ref{Experiment Result} discusses the results of experiment. Finally,
Section \ref{Lesson Learned} addresses the lesson learned from our experiment and potential future works.

\section{Proposed Architecture}  \label{Proposed Architecture}

\begin{figure*}
    \centering
    \includegraphics[width=0.7\linewidth]{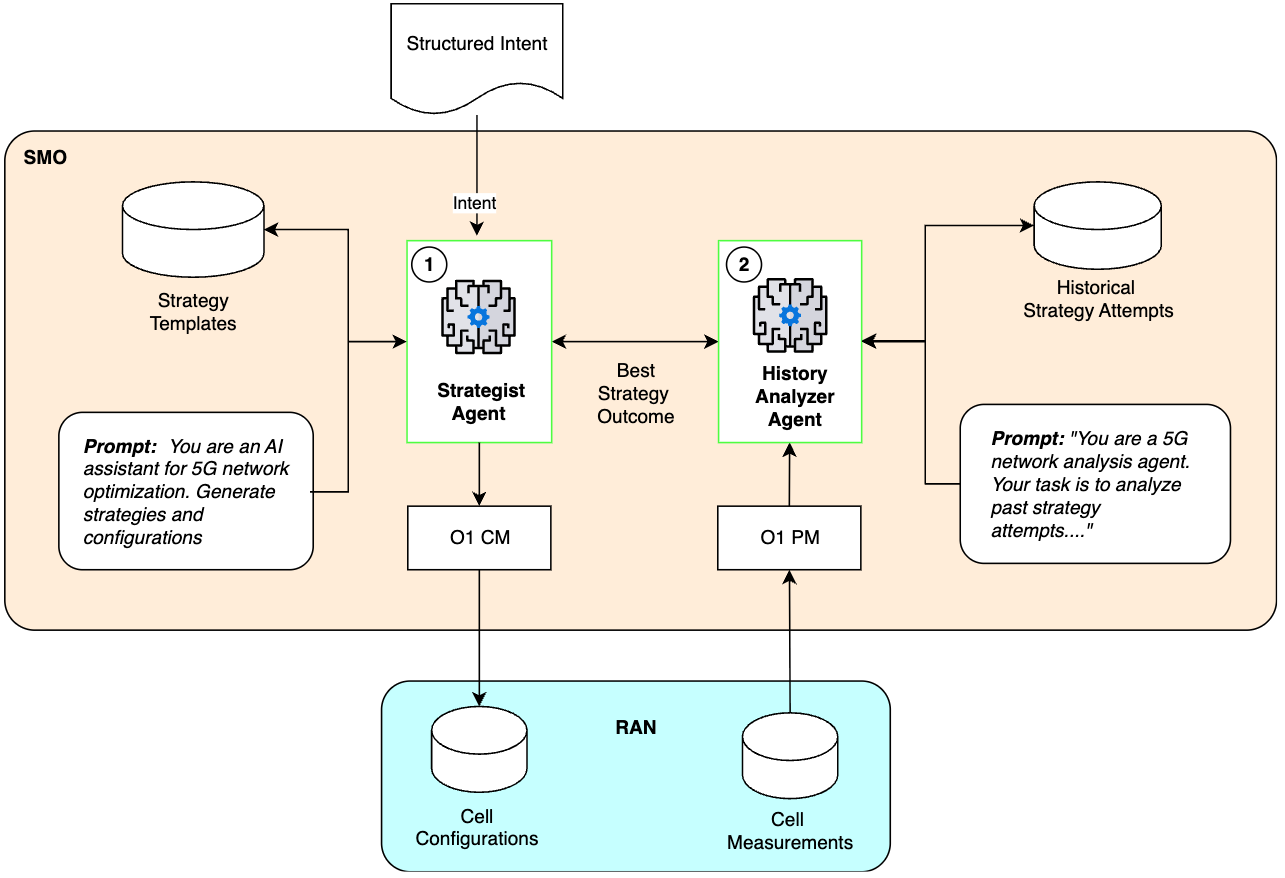}
    \caption{System architecture of an intent-based wireless network management system}
    \label{fig:intent-system}
\end{figure*}

This paper presents an intent-based wireless network management system, as shown in Fig. \ref{fig:intent-system}.
This system leverages cell configurations and cell measurement functions provided by the O-RAN experimental platform\cite{10464635}. Crucially, the interface establishing the connection between our system and the RAN adheres to O-RAN specifications. 

Our IBN system uses an agentic approach. A Strategist Agent, responsible for intent translation, receives the formalized intent declared by the consumer. This agent leverages a LLM to perform the translation, using the user prompt structure shown in Fig. \ref{fig:prompt-fig}.

Agentic LLMs serve as the intelligent engine that translate the structured intents and generate configuration strategy. This involves a multi-stage process. First, the agent analyzes the intent, understanding the target objects, the desired metric, the target condition, and the target value. Then agent then decomposes this objective into a series of actionable strategies. For planning and orchestration, the agentic LLM, orchestrated through a LangGraph flow\cite{langgraph}, determines the necessary steps to achieve the intent. This may involve interacting with the RAN simulator via O-RAN O1 interface to retrieve relevant cell details and performance metrics. The "Strategist Agent," as shown in Fig. \ref{fig:intent-system}, plays a key role in formulating a configuration strategy based on the interpreted intent and potentially leveraging past successful attempts stored by the "History Analyzer Agent." The "Orchestrator Agent" then takes charge of executing these strategies by calling the appropriate configuration tools via the O-RAN O1 interface to configure the RAN simulator. While the focus of closed-loop monitoring might lean towards deterministic approaches, the initial configuration and optimization driven by the agentic LLM set the stage for efficient network operation.

% need more lines to describe the two data. 
\subsection{Formalized Intent Structure}
\begin{figure}
    \centering
    \includegraphics[width=0.9\linewidth]{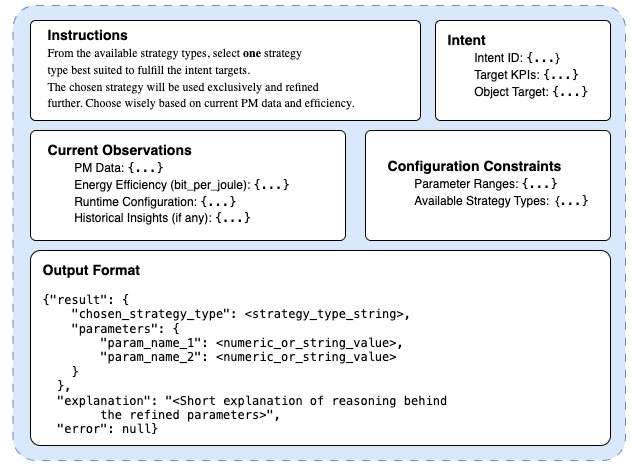}
    \caption{User prompt structure for LLM agent}
    \label{fig:prompt-fig}
\end{figure}

Declaration of intent by owner is formalized as a JSON object, referencing 3GPP TS 28.312\cite{28312}. This formalized intent is then parsed to populate the Intent context, which serves as a structured input for the system. As depicted in Fig. \ref{fig:prompt-fig}, the Intent context primarily comprises two distinct sections: Target KPIs and Object Target. The Target KPIs section specifies the performance metrics to be monitored and optimized (e.g., energy efficiency), along with their desired conditions and target values. Conversely, the Object Target section identifies the specific network elements or scopes to which the intent applies (e.g., a particular cell or sub-network). 
For instance, if a user wants to ensure that the RAN energy efficiency for Cell 7 within SubNetwork 1 is optimized to be as high as possible, specifically targeting a value greater than 800,000 bits per Joule, an example intent might specify the Object Target as follows:
\begin{lstlisting}[language=json,firstnumber=1]
{
    "objectInstance": "SubNetwork_1", 
    "ObjectTarget": ["Cell_7"]
}
\end{lstlisting}

And it declared alongside Target KPIs:

\begin{lstlisting}[language=json,firstnumber=1]
{
    [{
        "targetName": "RANEnergyEfficiency", 
         "targetCondition": "IS_GREATER_THAN", 
         "targetValue": "800000", 
         "targetUnit": "bit/joule"
     }] 
}
\end{lstlisting}

This comprehensive and structured representation of the intent serves as the initial input for our agentic LLM.

\subsection{Prompt Techniques}
We use one-shot prompting to effectively guide the LLM towards desired outputs. This technique provides a clear example of the expected output to ensure a consistent outcome. Specifically, we implement a five-section prompt structure to standardize input for LLM-driven intent translation. This structured approach is crucial for ensuring consistent and predictable LLM behavior, which is essential for the reliable and automated operation of an Intent-Based Network for RAN Management. Each section of the prompt is explicitly designed to provide comprehensive context and clear directives, allowing for precise control over the LLM's decision-making process. The distinct purposes of these sections are detailed below:

% We use prompt technique one-shot prompting to effectively guide the LLM towards desired outputs. We use five-section prompt structure to provide comprehensive context and clear directives. 
% Each prompt begins with an Instruction, explicitly stating the task the LLM is expected to perform. This is followed by the Intent section, which conveys the high-level objective or desired outcome from the system's perspective. To ensure that the LLM's decisions are grounded in real-world conditions, the Current Observation section provides real-time or relevant operational data. Crucially, Configuration Constraints are included to define any limitations, rules, or boundaries that the LLM must adhere to when formulating its response. Finally, the Output Format section specifies the exact structure in which the LLM's response should be delivered, ensuring parseability and seamless integration with downstream processes. This structured one-shot methodology enables precise control over the LLM's behavior while efficiently conveying all necessary information for accurate and constrained decision making.
\begin{itemize}
    \item \textbf{Instruction}: Each prompt begins with an Instruction, explicitly stating the task the LLM is expected to perform.
    \item \textbf{Intent}: Contain the high-level objective or desired outcome from the system's perspective.
    \item \textbf{Current Observation}: Provides real-time or relevant operational data to give context about the current state of the network.
    \item \textbf{Configuration Constraints}: This section is included to define any limitations for each RAN gNB, informing the LLM know what parameter and strategy are available to adjust.
    \item \textbf{Output Format}: Specifies the exact structure in which the LLM's response should be delivered. Defining structure output enables direct parsing for further computation processes.
\end{itemize}

\subsection{Agentic Framework}

Our system is built upon two distinct agents: the History Analyzer Agent and the Strategist Agent. The History Analyzer Agent is responsible for analyzing past strategies applied in the network and obtains real-time measurements directly from the RAN as input. Data about past strategy attempts are obtained from the Historical Strategy Attempts database. The History Analyzer Agent recommends the best strategy outcomes to the Strategist Agent. This processed data, along with a number of the most relevant previous strategy attempts, is then passed to the Strategist Agent for decision-making. Predefined strategy that contains data about target cell capabilities stored in the Strategy Templates database. Both agents perform their respective inference tasks calling the NVIDIA NIM inference API with detail inference as summarized in Table \ref{tab:hyperparameters}. The underlying intelligence for these agents is configured within the Prompt Processor component, which utilizes the LangGraph library \footnote{https://github.com/langchain-ai/langgraph}.

\begin{table}[h!]
    \centering
    \caption{LLM Inference Parameters}
    \label{tab:hyperparameters}
    \begin{tabular}{ cc }
        \toprule % This will be a bold/thick line at the top
        \textbf{Parameter} & \textbf{Value} \\
        \midrule % This will be a standard mid-rule
        Model & llama-3.1-70b-instruct \\
        Temperature & 0.2 \\
        Top P & 0.7 \\
        Max Tokens & 1024 \\
        Stream & True \\
        \bottomrule % This will be a bold/thick line at the bottom
    \end{tabular}
\end{table}

\subsection{O1 Configuration \& Performance Management}
To support the agentic architecture described in our Intent-Based Network for RAN Management, we leverage both Configuration Management (CM) and Performance Management (PM) over the O-RAN O1 interface to enable closed-loop network control. The agents continuously retrieve real-time network PM data which serve as critical input to assess the current operational state and evaluate the impact of previous configurations. For configuration, CM operations are performed using the NETCONF protocol. The agents first perform \verb|<get-config>| operations to retrieve current RAN configuration parameters (e.g., transmit power levels). Using interpreted user intent and historical performance context, the Strategist Agent subsequently issues \verb|<edit-config>| operations to apply optimized configurations to the RAN environment. This integration of CM and PM forms the backbone of the intent-based closed-loop control system.

This paper presents an intent-based wireless network management system, as shown in Fig. \ref{fig:intent-system}. 
% This system leverages cell configurations and cell measurements functions provided by the O-RAN experimental platform.

% \section{Methodology}
% \begin{table}[h!]
%     \centering
%     \caption{LLM Inference Parameters}
%     \label{tab:hyperparameters}
%     \begin{tabular}{ cc }
%         \toprule % This will be a bold/thick line at the top
%         \textbf{Parameter} & \textbf{Value} \\
%         \midrule % This will be a standard mid-rule
%         Model & llama-3.1-70b-instruct \\
%         Temperature & 0.2 \\
%         Top P & 0.7 \\
%         Max Tokens & 1024 \\
%         Stream & True \\
%         \bottomrule % This will be a bold/thick line at the bottom
%     \end{tabular}
% \end{table}

% Table \ref{tab:hyperparameters} summarizes the key hyperparameters and configurations.

\section{Experimental Result}  \label{Experiment Result}

This section presents the results of our experimental analysis to validate the effectiveness of the proposed LLM-based RAN management approach. The experimental environment was established using the VIAVI Solutions AI RSG RAN Simulator, which was configured with the parameters listed in Table \ref{tab:airsg}.

% table 3?
\begin{table}[h!]
    \centering
    \caption{Scenario Simulation}
    \label{tab:airsg}
    \begin{tabular}{ cc }
        \toprule % This will be a bold/thick line at the top
        \textbf{Parameter} & \textbf{Value} \\
        \midrule % This will be a standard mid-rule
        Simulator & VIAVI AI RSG \\
        Version & 2.4 \\
        Number of gNB & 2 \\
        Number of UE(s) at gNB$_1$ & 10 \\
        Number of UE(s) at gNB$_2$ & 8 \\
        Mobility Model & short range trajectory \\
        Movement speed & 0 \\
        \bottomrule % This will be a bold/thick line at the bottom
    \end{tabular}
\end{table}

\subsection{Single-Run Evaluation}

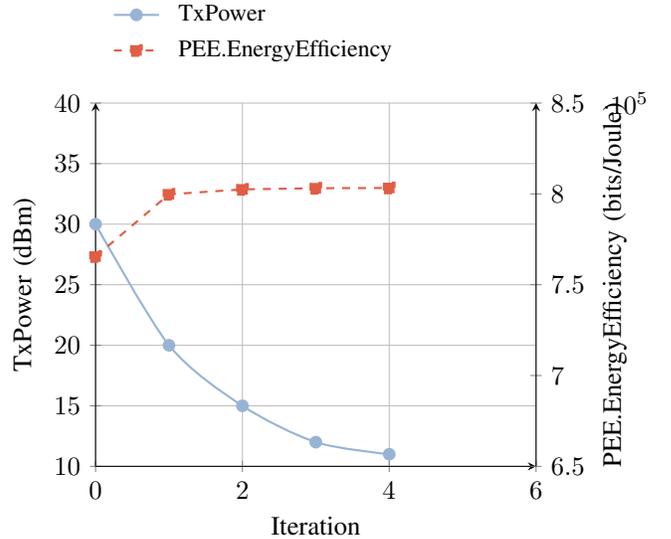
\begin{figure}[htbp!]
    \centering
    \begin{tikzpicture}
        % First Axis: Configured Tx Power (Left Y-axis)
        \begin{axis}[
            width=0.85\linewidth,
            xlabel={Iteration},
            ylabel={TxPower (dBm)},
            xmin=0, xmax=6, % Iteration range
            ymin=10, ymax=40, % Tx Power range
            ytick={10,15,20,25,30,35,40, 45}, % Specific ticks for Tx Power
            grid=major,
            axis x line=bottom, % Ensure x-axis is at the bottom
            axis y line=left,  % Primary Y-axis on the left
            % *** Legend style for the combined legend ***
            legend style={
                at={(0.02, 1.3)}, % Position: very close to top-left corner
                anchor=north west, % Anchor the top-left of the legend box at (0.02, 0.98)
                font=\small,      % Adjust font size as needed
                draw=none,        % No border around the legend (optional, IG1230 often cleaner)
                fill=none,        % No background fill (essential if you have overlapping elements)
                % Ensure entries are stacked vertically in a single column
                legend cell align=left, % Align legend text to the left
                column sep=5pt, % Space between columns if you were to have multiple
                row sep=2pt, % Space between legend rows
            },
            mark options={scale=0.8} % Styling for markers
        ]
            % Plot for Configured Tx Power
            \addplot[success, thick, smooth, mark=*, mark options={fill=success}] coordinates {
                (0, 30) (1, 20) (2, 15) (3, 12) (4, 11)
            };
            \addlegendentry{TxPower} % First legend entry

            % Add the legend image and entry for the second plot here
            % This explicitly tells the legend in *this* axis about the second plot.
            \addlegendimage{fail, thick, dashed, mark=square*, mark options={fill=fail}}
            \addlegendentry{PEE.EnergyEfficiency} % Second legend entry

        \end{axis}

        % Second Axis: PEE.EnergyEfficiency (Right Y-axis)
        % This axis only defines the plot and its right Y-axis properties.
        \begin{axis}[
            width=0.85\linewidth,
            xmin=0, xmax=6, % IMPORTANT: Keep xmin/xmax consistent with the first axis
            ymin=650000, ymax=850000, % Adjusted ymin/ymax to match ytick values
            ytick={650000,700000,750000,800000, 850000}, % Specific ticks for Energy Efficiency
            ylabel={PEE.EnergyEfficiency (bits/Joule)},
            axis x line=none,  % No x-axis line for this overlaid axis
            axis y line=right, % This Y-axis on the right
            xticklabel=\empty, % Remove x-labels
            xtick=\empty,      % No x-ticks
            enlargelimits=false, % Ensure it aligns perfectly with the first axis
            % *** IMPORTANT: NO legend style or \addlegendentry here! ***
            mark options={scale=0.5} % Styling for markers
        ]
            % Dummy Data for PEE.EnergyEfficiency (e.g., bits/Joule)
            \addplot[fail, thick, dashed, mark=square*, mark options={fill=fail}] coordinates {
                (0, 765493) (1, 799756) (2, 802493) (3, 803127) (4, 803257)
            };
            % No \addlegendentry here because we added it in the first axis
        \end{axis}
    \end{tikzpicture}
    \caption{Correlation between configured TxPower and PEE.EnergyEfficiency over Iteration(s) in a single-run test.}
    \label{fig:res1}
\end{figure}

Figure \ref{fig:res1} presents the correlation between configured \texttt{"TxPower"} and \texttt{"PEE.EnergyEfficiency"} for iteration(s). The result obtained by performing a close-loop iteration of our system. Each iteration performed by the system by refining strategies from previous state as reference towards the required KPI set by intent. As iterations progress, there is a clear trend of the \texttt{"TxPower"} (blue line, left y-axis) decreasing significantly, starting from approximately 30 dBm at iteration 0 and converging towards 11 dBm by iteration 4. In contrast, \texttt{"PEE.EnergyEfficiency"} (red dashed line, right y-axis) shows a corresponding increase, rising from around \SI{7.7e5}{bits/joule} at iteration 0 and stabilizing near \SI{8e5}{bits/joule} from iteration 2 onward. This inverse relationship suggests that the LLM's dynamic configuration of lower transmit power levels leads to an improved overall energy efficiency of the system, indicating a successful optimization process.

\subsection{Closed-Loop Precision Evaluation}

% We did 6 attempts with variation of target from 800, 803, 805. For this test we observe the behaiour system on to upper bound of energy efficiency.  

% We observe the different in number of iteration happen due to LLM response on the suggested value. The lower iteration happended due to LLM response more aggresive value towards the target. 
% We also observe that Fail rsutl at 7th attempt due to maximum value of parameters allowed to be adjusted achoeved thje max value. Therefore LLM concoude there is a plateau and decided that optimal value was reached at the previous configuration value and then abandon the strategy,

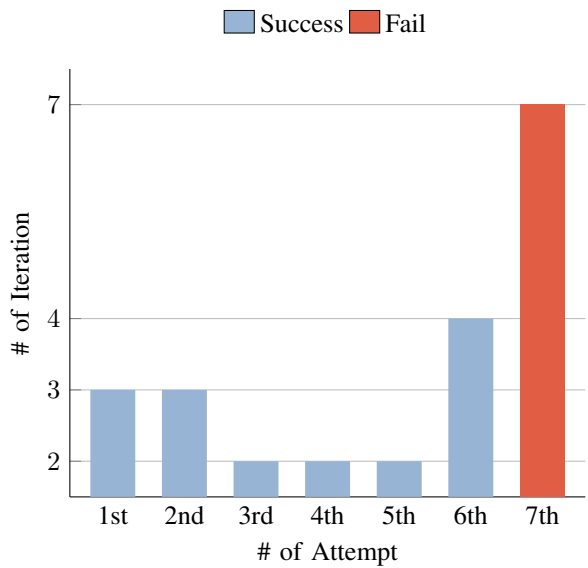
\begin{figure}[htbp!]
    \centering
\begin{tikzpicture}
    \begin{axis}[
        symbolic x coords={1st,2nd,3rd,4th,5th,6th,7th},
        xtick={1st,2nd,3rd,4th,5th,6th,7th},
        ytick={0,1,2,3,4,7},
        xlabel={\# of Attempt},
        ylabel={\# of Iteration},
        axis x line*=bottom,
        axis y line*=left,
        ymajorgrids,
        bar width=17pt,
        legend style={draw=none, at={(0.5,1.05)}, anchor=south, legend columns=-1},
        nodes near coords align={vertical},
        legend image code/.code={
            \draw[#1] (0cm,-0.1cm) rectangle (0.4cm,0.2cm);
        }
        ]

        % Success bars (combined into one \addplot)
        \addplot[ybar,draw=none,fill=success] coordinates {(1st,3) (2nd,3) (3rd,2) (4th,2) (5th,2) (6th,4)};
        \addlegendentry{Success}
        % Fail bar
        \addplot[ybar,draw=none,fill=fail] coordinates {(7th,7)};
        \addlegendentry{Fail}
        
    \end{axis}
\end{tikzpicture}
\caption{Closed-loop precision evaluation for an IBN System}
\label{close-loop-test}
\end{figure}

Figure \ref{close-loop-test} illustrates the experimental results of a closed-loop precision evaluation for an intent-based system, focusing on the system's performance across seven attempts with varying target values. The x-axis represents the number of attempts, ranging from 1st to 7th, while the y-axis indicates the number of iterations taken to achieve the target. An iteration indicates a cycle where the system performs a configuration, waits for the next PM sent by the target gNB, and completes the cycle upon receiving the PM. The first five attempts correspond to target value of 800. Then at 6th attempt we increase to 803 and finally is 805. The figures shows successful convergence with iteration counts of 3, 3, 2, 2, 2, and 4, respectively, reflecting the system's ability to adapt to different precision levels.

The variation in iteration counts highlights the influence of the LLM responses on the system's efficiency. Lower iteration numbers, such as 2, are observed when the LLM suggests more aggressive adjustments toward the target. In contrast, a higher iteration count of 4 at the sixth attempt suggests a more conservative adjustment process. The seventh attempt, marked by a "Fail" result with 7 iterations, indicates that the maximum allowable value of \texttt{txPower} to be adjusted was reached, leading the LLM to conclude a plateau. As a result, the system abandoned further optimization, as it was unable to fulfill the intent.

\subsection{Intent Boundary Testing}
Table \ref{boundary} presents the results of the Intent Boundary Testing, which evaluates the system's performance relative to the target \texttt{PEE.EnergyEfficiency}. The baseline value was established by recording initial simulation data using the VIAVI AI RSG prior to implementing any configurations. Subsequently, values both above and below this baseline were tested to assess system behavior. The results indicate that the system can process targets below the baseline, such as 750 kWh, by performing only monitoring and verification tasks, confirming that the system is already in a preferred state as defined by the intent. In contrast, when target  \texttt{PEE.EnergyEfficiency} set above baseline threshold starting at 803 kWh, the system attempts to fulfill the target until it reaches a plateau, beyond which no further optimization is possible, leading to abandonment of the process. This behavior is also corroborated by the findings in Fig. \ref{close-loop-test}.

\begin{table}[h!]
    \centering
    \caption{Intent Boundary Testing for \texttt{PEE.EnergyEfficiency}}
    \begin{tabular}{ c | c | r }
        \hline
        PEE.EnergyEfficiency (KWh) & Description & Result \\
        \hline
        765.4 & Baseline & \cellcolor{black!20}- \\
         >803 & Above Baseline Threshold & \cellcolor{red!20}Failed \\
        750 & Below Baseline Threshold & \cellcolor{blue!20}Passed \\
        % 770-803 & Within Baseline Range & \cellcolor{blue!20}Passed \\
        \hline  
    \end{tabular}
    \label{boundary}
\end{table}

\section{Lesson Learned and Future Works} \label{Lesson Learned}

This study presents a novel Intent-Based Network for RAN Management that harnesses LLMs to optimize the management of RAN configurations. Our primary contribution lies in developing a formalized prompt engineering technique that translates high-level user intents into a structured JSON format, encapsulating intent, network topology, KPIs, and configuration parameters, thus providing a standardized interface for LLMs to interpret and achieve the desired network objectives. We further introduce an agentic structure where two LLM agents collaboratively refine network configurations through iterative optimization in a close-loop manner.
This iterative process ensures the intent is achieved through gradual parameter adjustments based on real-time feedback thereby preventing network disruptions caused by abrupt configuration changes.

However, during our preliminary evaluations, we identified two key limitations that critically impact the reliability and efficacy of LLM-driven network management: challenges related to contextual understanding and data accuracy.

% the framework’s sensitivity to inaccuracies in Performance Monitoring (PM) data, which can lead to suboptimal configurations. This challenge underscores the need for future research into reinforcement learning (RL) algorithms to achieve system resilience. RL can adapt to noisy or outlier data via iterative feedback and reward-based optimization. Pursuing this research direction will advance the development of robust and intelligent 5G RAN management, ensuring consistent performance under diverse and imperfect data conditions.

\subsection{Contextual Understanding and Domain Specificity}
During the preliminary evaluation of our system, we observed that feeding comprehensive PM data directly to the LLM sometimes led to hallucinations or inconsistent interpretations in the generated configuration strategy. This highlights a critical lesson: while LLMs possess remarkable capabilities in natural language understanding, their ability to generate accurate and reliable responses within highly specialized domains, such as RAN configuration, is significantly enhanced by providing them with domain-specific ground truth. 

This challenge highlights the need for future research to use a structured approach to serve data, explicitly detailing the relationships between entities. Such ground truth, ideally represented through formalisms like RDF (Resource Description Framework) or ontology, contains the precise nature of network entities and the fundamental properties of parameters (e.g., the exact relationship between a configuration parameter and a measurement metric). Representing and leveraging this complex domain knowledge before it is fed into the LLM is crucial for generating accurate, concise, and scientifically sound 'justification' texts that the LLM can translate effectively. 

\subsection{The Impact of PM Data Inaccuracy}

The operational efficacy of LLMs within a network management framework is predicated on its direct interaction with, and inherent assumption that, PM data represents the 'ground truth.' A significant vulnerability arises from this unquestioning reliance, as our experiments revealed a critical aggregation issue. Although the RAN simulator transmits PM metrics at fixed intervals, the aggregated values often do not accurately reflect the network's true state due to the various aggregation algorithms and configurable time windows.

When PM data is inaccurate, it sets off a series of consequences. A flawed initial understanding leads the LLM to make suboptimal decisions, which in turn generate new, equally unreliable PM data. This feedback loop of misinformation can rapidly destabilize the network, ultimately undermining the purpose of the automated management system.

Therefore, future research must address this challenge by exploring alternative data collection methodologies. One promising solution is to use PM metrics as a streaming data source without aggregation at the gNB to ensure all data is treated as a real-time reflection of the network state. Furthermore, integrating reinforcement learning (RL) algorithms can enhance system resilience by allowing the system to adapt to noisy or outlier data via iterative feedback and reward-based optimization. This will advance the development of robust and intelligent RAN management.
% \section*{Acknowledgment}

% The preferred spelling of the word ``acknowledgment'' in America is without 
% an ``e'' after the ``g''. Avoid the stilted expression ``one of us (R. B. 
% G.) thanks $\ldots$''. Instead, try ``R. B. G. thanks$\ldots$''. Put sponsor 
% acknowledgments in the unnumbered footnote on the first page.

\section*{Acknowledgment}
This work was supported by the National Science and
Technology Council, Taiwan, under Contract numbers 111-2221-E-011-069-MY3, 112-2218-E-011-004, and 112-2218-E-011-006.

\bibliographystyle{IEEEtran}
\bibliography{bib}

\vspace{12pt}

\end{document}